\documentclass[onecolumn]{IEEEtran}

\usepackage[cmex10]{amsmath}
\usepackage{amssymb,bbm}
\usepackage{graphicx,color}

\newtheorem{lemma}{Lemma}[section]
\newtheorem{proposition}{Proposition}[section]
\newtheorem{theorem}{Theorem}[section]

\newtheorem{definition}{Definition}[section]

\def\endproof{{\hfill $\clubsuit$ \medskip}}

\def\clip{\mbox{Clip}}

\title{Estimation of Bandlimited Signals in Additive Gaussian Noise:
a ``Precision Indifference'' Principle}

\author{Animesh Kumar$^1$ and Vinod M.~Prabhakaran$^2$\\
$^1$ Department of Electrical Engineering, Indian Institute of Technology
Bombay, India \\
$^2$ School of Technology and Computer Science, Tata Institute of Fundamental
Research, Bombay, India \\
Emails: {animesh@ee.iitb.ac.in, vinodmp@tifr.res.in} }

\def\ltwo{{\cal L}^2}

\def\linf{{\cal L}^\infty}
\def\sinc{\mbox{sinc}}

\def\indicator{{\mathbbm{1}}}

\def\BLBounded{{\cal S}_{\mbox{\footnotesize BL,bdd}}}
\def\BLInterest{{BL}_{\mbox{\footnotesize int}}}

\def\onebit{\mbox{\footnotesize 1-bit}}
\def\frame{\mbox{\footnotesize fr}}
\def\rec{\mbox{\footnotesize rec}}
\def\var{\mbox{var}}
\def\conv{\star}

\def\pP{{\mathbb P}}

\def\eE{{\mathbb E}}
\def\rR{{\mathbb R}}
\def\zZ{{\mathbb Z}}

\begin{document}

\maketitle
\begin{abstract}
The sampling, quantization, and estimation of a bounded dynamic-range
bandlimited signal affected by additive independent Gaussian noise is
studied in this work.  For bandlimited signals, the distortion due to
additive independent Gaussian noise can be reduced by oversampling
(statistical diversity).  The pointwise expected mean-squared error is used
as a distortion metric for signal estimate in this work.  Two extreme
scenarios of quantizer precision are considered: (i) infinite precision
(real scalars); and (ii) one-bit quantization (sign information).  If $N$
is the oversampling ratio with respect to the Nyquist rate, then the
optimal law for distortion is $O(1/N)$.  We show that a distortion of
$O(1/N)$ can be achieved \textit{irrespective of the quantizer precision}
by considering the above-mentioned two extreme scenarios of quantization.
Thus, a \textit{quantization precision indifference} principle is
discovered, where the reconstruction distortion law, up to a
proportionality constant, is unaffected by quantizer's accuracy.
\end{abstract}
\begin{keywords}
bandlimited signals, sampling, estimation, quantization
\end{keywords}
\section{Introduction}
\label{sec:introduction}

Consider a bandlimited signal (or field) quantization problem, where the
samples are affected by additive independent and identically distributed
(i.i.d.) Gaussian noise.  For example, a spatial signal affected by
additive i.i.d.~Gaussian noise has to be sampled using an array of sensors.
In a distributed setup, where filtering before sampling is not possible,
noise in bandlimited signals can be reduced by statistical averaging of
independent noisy samples.  In addition, quantization error can be reduced
by oversampling as well as by increasing the analog-to-digital converter
(ADC) or quantizer precision.  The \textit{fundamental tradeoff} between
oversampling, quantizer precision, and (statistical) average distortion is
of interest. 

The tradeoffs between all subsets of these three quantities has been
studied in the literature. Tradeoffs for average distortion and
oversampling, additive Gaussian noise and average distortion with
unquantized samples, and oversampling and quantization have a flurry of
work (e.g.,
see~\cite{mallatSA2009,pinskerO1980,grayo1987,zoranDLS2007,kumarIRH2011}).
In this work, the tradeoff between oversampling, quantizer precision, and
the average distortion is of interest.

If extremely high-precision ADCs are used, then the sample distortion is
noise limited. On the other hand, if lowest precision single-bit ADCs are
used, then the sample distortion is limited by quantization. At a
high-level, it is expected that the distortion optimal ADC precision should
be `in between' these two extreme cases, in the sense that it should be
able to resolve the signal up to the noise level. Contrary to this
intuition, in this work it is shown that a distortion inversely
proportional to the oversampling above the Nyquist rate is
\textit{achievable} with single-bit quantizers.  With unquantized (infinite
precision) samples, the optimal distortion is \textit{speculated} to be
inversely proportional to the oversampling above the Nyquist rate in the
presence of independent Gaussian noise.  Accordingly, the focus of this
work is on the quantization of an additive independent Gaussian noise
affected bandlimited signal using single-bit ADCs and oversampling. The key
result of this paper is the uncovering of a quantization \textit{precision
indifference} principle, which is stated next.

\textit{Precision indifference principle:} Consider a bounded-dynamic range
bandlimited signal with samples affected by additive independent Gaussian
noise and observed through quantizers. If $N$ is the oversampling ratio,
with respect to the Nyquist rate, then the optimal law for maximum
pointwise mean-squared error is $O(1/N)$, irrespective of the quantizer
precision. In other words, for large $N$, the quantizer precision only
affects the proportionality constant of the distortion.

\textit{Prior art:} Averaging and other properties of independent random
variables are well studied in statistics~\cite{bickelDM2001}. Quantization
error can be reduced by oversampling as well as by increasing the ADC
(quantizer)
precision~(see~\cite{grayo1987,GrayNIT98,zoranDLS2007,kumarIRH2011} for the
entire range of results).  Estimation of square-integrable signals in the
presence of Gaussian noise was studied by Pinsker~\cite{pinskerO1980};
however, quantization is not addressed in his work.  Signal quantization
with additive noise as a dither has been studied by
Masry~\cite{masryT1981}, however signal was not assumed to be bandlimited.
Masry's results give a decay of $O(1/N^{2/3})$ for bandlimited signals,
where $N$ is the oversampling above the Nyquist rate; this decay is slower
than an $O(1/N)$ decay that we are after.  The sampling of signals defined
on a finite support, while using single-bit quantizers in the presence of
ambient noise, has been also studied \cite{wangID2009,masryIF2009}.

\textit{Notation:} The set of bounded signals and the set of finite energy
signals  will be denoted by $\linf(\rR)$ and $\ltwo(\rR)$, respectively.
The signal of interest will be denoted by $g(t)$.  For a signal $s(t)$ in
$\ltwo(\rR)$, the Fourier transform will be denoted by $\tilde{s}(\omega)$.
The Fourier transform and its inverse are defined as,
\begin{align}
\tilde{s}(\omega) = \int_{\rR} s(t) \exp(- j\omega t)\mbox{d}t; \ \ s(t) =
\frac{1}{2\pi} \int_{\rR} \tilde{s}(\omega) \exp(j \omega t)
\mbox{d}\omega.  \nonumber
\end{align}
The indicator function of a set $A$ is denoted by $\indicator(x \in A)$.
Random variables or processes will be denoted by uppercase letters. The
additive independent Gaussian noise is denoted by $W(t)$. The set of reals
and integers will be denoted by $\rR$ and $\zZ$, respectively. The
cumulative distribution function (cdf) of a Gaussian random variable with
mean zero and variance $\sigma^2$ will be denoted by $F(x), x\in \rR$.  The
convolution and expectation operation will be denoted by $\conv$ and $\eE$,
respectively. It is assumed that all probability models have an underlying
sample space, sigma-field, and probability measure such that (weighted)
averages and indicator functions are measurable.

\textit{Organization:} The mathematical formulation of our sampling problem
is discussed in Sec.~\ref{sec:formulation}. Short review of stable
interpolation kernels and smoothness properties of associated signals are
discussed in Sec.~\ref{sec:background}. The discussion on precision
indifference principle appears in Sec.~\ref{sec:estimation}. Estimation
with perfect samples and single-bit quantized samples are discussed in
Sec.~\ref{sec:perfectsamples} and Sec.~\ref{sec:onebit}, respectively.
Conclusions are presented in Sec.~\ref{sec:conclusions}. To maintain the
flow of the paper, long proofs appear in the Appendix.

\section{Problem formulation}
\label{sec:formulation}

\begin{figure}[!htb]
\begin{center} \scalebox{1.0}{\input{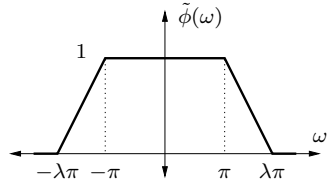}} 
\end{center}
\caption{\label{fig:stablesincsquare} \sl \small \textbf{Stable
interpolation filter:} The kernel $\phi(t) \leftrightarrow
\tilde{\phi}(\omega)$ is defined in (\ref{eq:phit}); this kernel is
absolutely integrable and will be used to define bounded bandlimited
signals.}
\end{figure}

The discussion begins with a quick review of a stable bandlimited kernel,
which is essential for stable interpolation in $\linf(\rR)$ and for
defining bandlimited signals.  For $\lambda > 1$ and $a = (\lambda - 1)/2$,
consider the kernel $\phi (t)$ that is given by
\begin{align}
\phi(t) =  \frac{1}{\pi a t^2} \sin ((\pi + a) t) \sin (a t); \ \ \phi(0) =
1+ \frac{a}{\pi}. \label{eq:phit}
\end{align}
The kernel decreases sufficiently fast (approximately as $1/t^2$) and
therefore it is absolutely and square integrable. Its Fourier transform is
illustrated in Fig.~\ref{fig:stablesincsquare}.
This kernel can be used to define the set of bounded bandlimited signals,
which is a subset of the Zakai class of bandlimited
signals~\cite{zakaiB1965}.  Consider
\begin{align}
\BLInterest := \{g(t) : |g(t)| \leq 1 \mbox{ and } g(t) \conv \phi(t) =
g(t) \ \forall  t \in \rR\}. \label{eq:blinterest}
\end{align}
The above definition ensures that $g(t)$ is continuous everywhere. It is
easy to verify that the set of bounded bandlimited signals in $\ltwo(\rR)$
with Fourier spectrum zero outside $[-\pi, \pi]$ also belongs to the set
$\BLInterest$. The set $\BLInterest$ also includes (almost-surely) any
sample path of a bounded-dynamic range bandlimited wide-sense stationary
process~\cite{cambanisMZ1976}. The quantization of bandlimited signals from
the set $\BLInterest$ in the presence of additive independent Gaussian
noise is studied in this work. \textit{The derived results are applicable
to finite energy bounded bandlimited signals as well as (almost surely) to
any sample path of a bounded wide-sense stationary bandlimited process.}

The signal affected by additive noise, $g(t) + W(t)$, is available for
sampling.  It is assumed that $W(t) \sim {\cal N}(0, \sigma^2)$ for all
$t\in\rR$.  Independence of noise implies that $W(t_1), W(t_2), \ldots,
W(t_n)$ for distinct $t_1, t_2, \ldots, t_n \in \rR$ are i.i.d.~with ${\cal
N}(0, \sigma^2)$ distribution. The Nyquist rate at which $g(t)$ should be
sampled for perfect reconstruction is one sample/second.  In the noise-free
regime, when $\sigma = 0$, it is sufficient to sample $g(t)$ at the Nyquist
rate for convergence in $\linf(\rR)$.  In the noise-limited regime, when
$\sigma > 0$, the reconstruction based on samples of $g(t)$ will have
distortion (statistical mean-squared error).  This distortion can be
reduced by oversampling. Let $N$, a positive integer, be the oversampling
rate.  For any statistical estimate $\widehat{G}_{\rec}(t)$ of the signal
$g(t)$, the maximum pointwise mean-squared error $D_{\rec}$ is defined as
the \textit{distortion}, i.e.,
\begin{align}
D_{\rec} := \sup_{t\in\rR} D_{\rec}(t) = \sup_{t\in\rR} \eE\left|
\widehat{G}_{\rec}(t) - g(t) \right|^2. \label{eq:distortion}
\end{align}
For a  pointwise-consistent reconstruction, the distortion in
(\ref{eq:distortion}) should decrease to zero as the oversampling rate $N$
increases to infinity~\cite{bickelDM2001}.  Consistent reconstruction of
smooth signals with a random dither, in the presence of single-bit
quantizers, has been obtained in the past~\cite{masryT1981}; therefore, the
asymptotic rate of decrease in $D_{\rec}$ with $N$ is of interest to us.
Due to finite precision limitations (ADC operation) during acquisition, the
signal samples are quantized.  Since quantization is a lossy
operation~\cite{gershogray}, $D_{\rec}$ is expected to depend upon the ADC
precision employed. As mentioned in Sec.~\ref{sec:introduction}, it will be
shown that $D_{\rec}$ decreases as $O(1/N)$, irrespective of the sensor
precision. Thus, the ADC precision only manifests in the proportionality
constant (independent of the oversampling factor $N$) in the optimal
asymptotic reconstruction distortion.

\begin{figure}[!htb]
\begin{center} \scalebox{1.0}{\input{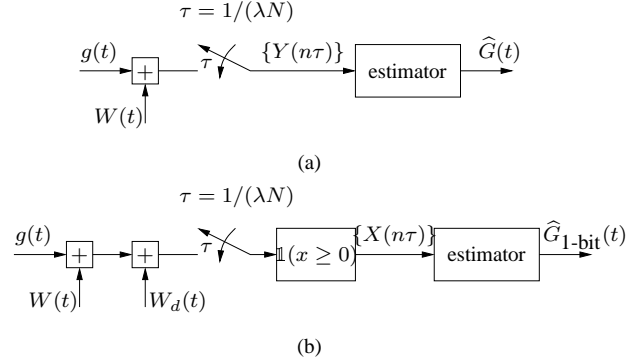}} 
\end{center}
\caption{\label{fig:bitdestructionblocks} \sl \small \textbf{Two extreme
scenarios of quantization:} In both the scenarios the signal $g(t)$ is
observed with additive independent Gaussian noise $W(t)$. In (a), the
estimator works with infinite precision (unquantized) samples $\{Y(n \tau),
n \in \zZ\}$. In (b), the estimator works with poorest precision (one-bit)
samples $\{X(n) , n \in \zZ\}$ where $X(n \tau) = \indicator(Y(n\tau) \geq
0)$.}
\end{figure}

To show the proposed precision indifference principle, two extreme cases of
quantizer precision will be analyzed and their distortion will be compared:
(i) signal distortion with perfect samples; and (ii) signal distortion with
samples quantized using single-bit ADCs.  The sampling setup for these two
cases are illustrated in Fig.~\ref{fig:bitdestructionblocks}.  In
Fig.~\ref{fig:bitdestructionblocks}(a), the estimator works with infinite
precision (unquantized) noisy samples while in
Fig.~\ref{fig:bitdestructionblocks}(b), the estimator works with poorest
precision (one-bit) noisy samples. The role of extra dither noise $W_d(t)$
will be explained later in Sec.~\ref{sec:onebit}.  The estimator
$\widehat{G}_{\onebit}(t)$ will be designed and its distortion performance
will be analyzed in this work.

Before we move on to the next section, it should be noted that the kernel
$\phi(t)$ and its derivative $\phi'(t)$ are absolutely integrable.  This
absolute integrability and square integrability of $\phi(t)$ can be
translated into the following observations, which will be useful in
Sec.~\ref{sec:onebit} during distortion analysis:
\begin{align}
C_{\phi} & := \int_{t \in \rR} |\phi(t)| \mbox{d} t < \infty,
\label{eq:integrablephi} \\
C_{\phi}' & := \sup_{\{t_k: t_k \in \left[k/\lambda, (k+1)/\lambda \right],
k \in \zZ\}} \sum_{k\in \zZ} |\phi'(t_k)| < \infty.
\label{eq:summablephiprime} \\
\mbox{and } C_{\phi}'' & := \sup_{t \in \rR}\sum_{k \in \zZ} \left| \phi
\left(t - \frac{k}{\lambda}\right) \right|^2 < \infty.
\label{eq:squaresummablephi}
\end{align}
The next section will review pertinent mathematical results which will be
used in the later sections.

\section{Background}
\label{sec:background}

The stable interpolation formula for Zakai sense bandlimited signals is
discussd first. The necessity of $W_d$ and associated variance-conditions
on Gaussian noise (see~Fig.~\ref{fig:bitdestructionblocks}(b)) are
discussed. If the pointwise error in  interpolation is bounded with bounded
perturbation of samples, it is called as a \textit{stable interpolation}.
The properties of stable interpolation and its implications on filtering of
bounded signals are given at the end of this section.

>From the interpolation formula for Zakai sense bandlimited signals, the
signal of interest $g(t)$ can be perfectly reconstructed from its samples
taken at the Nyquist rate. For $g(t) \in \BLInterest$, the interpolation
formula is given by~\cite[Lemma~3.1]{kumarIRH2011},
\begin{align}
g(t) = \lambda \sum_{n \in \zZ} g \left(\frac{n}{\lambda}\right) \phi
\left(t - \frac{n}{\lambda}\right) . \label{eq:sincinterp}
\end{align}
where the equality holds absolutely, pointwise, and in $\linf(\rR)$. Thus,
in the absence of noise, it is sufficient to sample $g(t)$ at a rate of
$\lambda$ sample per second (or per meter in the context of spatial
fields).  In the presence of quantization, the reconstruction in
(\ref{eq:sincinterp}) is stable in $\linf(\rR)$. 

The role of $W_d(t)$ in Fig.~\ref{fig:bitdestructionblocks}(b) will now be
highlighted. If the noise variance $\sigma$ is very small compared to the
dynamic range of the signal $g(t)$, i.e., $|\sigma| \ll 1$, then the
samples $\indicator( g(t) + W(t) \geq 0)$ will not capture small scale
local variation in $g(t)$.  Due to quantization, estimators such as maximum
likelihood are expected to be non-linear and their analysis is too complex.
To alleviate this issue, if $\var(W(t)) = \sigma^2$ is very small, then an
extra additive independent Gaussian dither $W_d(t)$ can be added to ensure
that $\indicator( g(t) + W(t) + W_d(t) \geq 0)$ is sufficiently random. It
is assumed that $W_d(t)$ and $W(t)$ are independent. Such dithering allows
us to use an analytically tractable reconstruction procedure, which has
\textit{order-optimal distortion}.  The block-diagram for sampling with
one-bit ADCs is illustrated in Fig.~\ref{fig:bitdestructionblocks}(b). The
technical condition on $\sigma^2 = \var(W(t)) + \var(W_d(t))$ is stated
using the cdf of $W+W_d$.  Let $F: \rR \rightarrow [0,1]$ be the cdf of
$W+W_d$.  Let $f(x)$ be the associated probability density function with
$f(\pm C_\phi) = \delta$ and $f(0) = \Delta$.  Observe that $\Delta >
\delta$, since $f(x) = \frac{1}{\sqrt{2\pi} \sigma} \exp( - x^2/2
\sigma^2)$.  It is required that there is a parameter $\mu > 0$ such that
\begin{align}
\left( 1 - \frac{1}{\sqrt{2} C^2_\phi} \right) \frac{1}{\delta} < \mu <
\frac{1}{\Delta}, \label{eq:variancecondition}
\end{align}
where $C_\phi$ is the constant in (\ref{eq:integrablephi}). First fix a
$\lambda > 1$.  Then, $C_\phi = \int_{t \in \rR} |\phi(t)| \mbox{d}t >
\int_{t \in \rR} \phi(t) \mbox{d} t = \tilde{\phi}(0) = 1$. That is,
$C^2_\phi \sqrt{2} > \sqrt{2} > 1$. Therefore, the lower bound on $\mu$ in
(\ref{eq:variancecondition}) is positive.  Next, observe that if $\sigma$
is large but fixed, then $\delta = f(C_\phi) \approx f(0) = \Delta$. Then
$\delta$ and $\Delta$ are close enough and the inequality in
(\ref{eq:variancecondition}) can be satisfied. In other words, for a fixed
$\lambda$ and hence $C_\phi$, there is a \textit{finite} number $\sigma_0$
for which (\ref{eq:variancecondition}) is satisfied for all $\sigma >
\sigma_0$. If $\var(W(t)) < \sigma_{0}^2$, then $\var(W_d(t)) >
\sigma_{0}^{2} - \var(W(t))$ will ensure that $\var(W + W_d) >
\sigma_{0}^2$. If $\var(W(t)) \geq \sigma_{0}^2$, then the extra dither is
not needed. This condition will be used in the distortion analysis in
Sec.~\ref{sec:onebit}.

For single-bit estimation, the signal $F(g(t)) - 1/2$ will be encountered,
where $F:\rR \rightarrow [0,1]$ is the cumulative distribution function of
the stationary noise random variable $W(t) + W_d(t)$.  Since $g(t) \in [-1,
1]$, and $F(x)$ has a wider support than the dynamic range of signal (i.e.,
$[-1, 1]$), therefore $F'(x)$ is finite and non-zero for $x \in [-1,1]$.
Since $F(0) = 1/2$ by symmetry, therefore, $F(g(t)) - 1/2$ is more
convenient than $F(g(t))$ to work with.  For simplicity of notation, let
$l(t) = F(g(t)) - 1/2$.  Then $|l(t)| \leq |F(1)| - 1/2$, i.e., $l(t)$ is
bounded.  The bound depends only on the noise distribution and the dynamic
range of $g(t)$.  Finally $|l'(t)| = |F'(g(t)) g'(t)| \leq |F'(0) 2 \pi^2|$
since $F'(0)$ maximizes $F'(x)$ in $[-1,1]$ and $|g'(t)|\leq 2 \pi^2$
(see~\cite[Proposition~3.1]{kumarIRH2011}).

The definition of $\BLInterest$ involves convolution with a stable
kernel and convolution will often appear in the context of error
analysis. The following short lemma will be quite useful later on.
\begin{lemma}% [Error propagation through $\phi(t)$ filter] 
\label{lemma:propagation}
Let $p(t)$ be a signal such that $||p||_\infty$ is finite and $P(t)$
be any random process such that $P(t)$ is bounded (i.e., $\sup_{t \in
\rR} \eE(P^2(t))$ is finite). Then,
\begin{align}
||p \star \phi ||_\infty & \leq C_\phi ||p||_\infty,
\label{eq:stableerror}\\
\mbox{and } \eE[( |P(t)| \star |\phi(t)|)^2] & \leq C^2_\phi \sup_{t
\in \rR} \eE(P^2(t)), \label{eq:stablevariance}
\end{align}
where the convolutions are well defined since $\phi(t)$ is absolutely
integrable.
\end{lemma}

\IEEEproof The proof follows by the definition of convolution and the
triangle inequality. We have
\begin{align}
|p(t) \star \phi(t)| & = \left| \int_{u \in \rR} p(u) \phi(t-u)
\mbox{d}u \right|, \nonumber \\
& \leq \int_{u \in \rR} |p(u)| |\phi(t-u)| \mbox{d}u, \nonumber \\
& \leq ||p||_\infty \int_{u \in \rR} |\phi(t - u)| \mbox{d}u,
\nonumber \\
& = C_\phi ||p||_\infty. \nonumber
\end{align}
For the second moment bound, note that
\begin{align}
& \eE(||P(t)| \star |\phi(t)||^2) \nonumber  \\
& = \eE\left( \iint_{u, v \in \rR} |P(u)| |P(v)| |\phi(t-u)
|\phi(t-v)| \mbox{d}u \mbox{d}v \right), \nonumber \\
& = \iint_{u, v \in \rR} \eE( |P(u)| |P(v)|) |\phi(t-u)| |\phi(t-v)|
\mbox{d}u \mbox{d}v ,\nonumber \\
& \stackrel{(a)}{\leq}  \sup_{t \in \rR} \eE(P^2(t)) \iint_{u, v \in
\rR} |\phi(t-u)| |\phi(t-v)| \mbox{d}u \mbox{d}v, \nonumber \\
& = C^2_\phi \sup_{t \in \rR} \eE(P^2(t)), \nonumber 
\end{align}
where $(a)$ follows by $\eE(2|P(u)| |P(v)|) \leq \eE(P^2(u) + P^2(v))
\leq 2 \sup_{t} \eE(P^2(t))$. Thus the proof is complete. \endproof

The two extreme scenarios of quantization as depicted in
Fig.~\ref{fig:bitdestructionblocks} and their distortions will now be
analyzed in the next section.

\section{Estimation of bandlimited signal}
\label{sec:estimation}

Interpolation of bandlimited signals with perfect samples is a well
known topic~\cite{mallatSA2009}. Loosely speaking, a bandlimited
signal of duration $T$ and bandwidth $\pi$ has $2\pi T$ degrees of
freedom~\cite{slepianO1976}.  With $N T$ noisy samples of the field
$g(t) + W(t)$  in duration $T$, the optimal distortion is expected to
be $O(1/N)$.\footnote{A single bounded constant in additive
independent Gaussian noise with $N$ independent readings can be
estimated up to a distortion of $O(1/N)$~\cite{bickelDM2001}.} With
this note, sampling schemes with oversampling rate $N$ are designed to
achieve a distortion of $O(1/N)$ for sampling $g(t)$.

\subsection{Estimation with perfect samples}
\label{sec:perfectsamples}

A brief review of estimation with perfect samples will be highlighted
first. Optimal minimum mean-squared method can be found in the work of
Pinsker~\cite{pinskerO1980}. For illustration and to get a distortion
proportional to $O(1/N)$, it suffices to use the frame expansion.  Let
the integer-valued oversampling ratio (above the Nyquist rate) be $N$,
and $\tau = 1/(\lambda N)$. Then, the samples $\{Y(n \tau), n \in \zZ
\}$ are available for the reconstruction of $g(t)$.  Using frame
expansion or the shift-invariance of bandlimited signals, 
\begin{eqnarray}
g(t) &=& \frac{1}{N}  \sum_{n \in \zZ} \lambda g(n \tau)\phi(t - n
\tau), \nonumber \\
& = & \frac{1}{N} \sum_{i = 0}^{N-1} \sum_{k \in \zZ} \lambda  g\left(
\frac{k}{\lambda} + i \tau \right) \phi\left(t - \frac{k}{\lambda} -
\frac{i}{N \lambda}\right) ,\label{eq:frameexpansion}
\end{eqnarray}
where the equality holds pointwise and in $\linf(\rR)$. It must be
noted that the basic operation in (\ref{eq:frameexpansion}) is that of
averaging; hence, the noise is expected to average out while the
signal will be retained.  This intuition motivates the following
estimator for $g(t)$ from noisy data (see
Fig.~\ref{fig:bitdestructionblocks}(a)). Define
\begin{align}
\widehat{G}_{\frame}(t) & := \frac{1}{N} \sum_{n \in \zZ} \lambda Y(n
\tau)\phi(t - n \tau), \label{eq:frameestimate}\\
& = \frac{1}{N} \sum_{i = 0}^{N-1} \sum_{k \in \zZ} \lambda [g+W]
\left( \frac{k}{\lambda} + i \tau \right) \phi\left(t -
\frac{k}{\lambda} - \frac{i}{N \lambda}\right).\nonumber
\end{align}
The distortion of $\widehat{G}_{\frame}(t)$ is given by the following
proposition.
\begin{proposition}[Frame estimate with $O(1/N)$ distortion]
\label{prop:mse_unquantized}
Let $\widehat{G}_{\frame}(t)$ in (\ref{eq:frameestimate}) be an
estimate for the bandlimited field $g(t)$ corrupted by additive
independent Gaussian noise. Let $D_{\frame}(t) := \eE |
\widehat{G}_{\frame}(t) - g(t)|^2$.  Then,
\begin{eqnarray}
\sup_{t \in \rR} D_{\frame}(t) \leq \frac{C_{\phi}'' \lambda^2
\sigma^2}{N}
\end{eqnarray}
where the constants $\sigma^2$ and $C_{\phi}''$ from
(\ref{eq:squaresummablephi}) do not depend on $N$. 
\end{proposition}

\IEEEproof See Appendix~\ref{ap:unquantizedMSE}. \endproof

The signal term in (\ref{eq:frameestimate}) converges in $\linf(\rR)$
to $g(t)$.  The noise term results in an independent sum of zero-mean
random variables at every $t \in \rR$.  This sum of random variables
has a variance that decreases as $(1/N)$ due to the finite energy of
the interpolation kernel $\phi(t)$. The constant $C_{\phi}''$ depends
on the properties of the kernel $\phi(t)$. The estimation with
single-bit quantizers and associated distortion analysis will be
presented next.

\subsection{Estimation with single-bit quantized samples}
\label{sec:onebit}

This section will present the key result of this work. Consider the system
illustrated in Fig.~\ref{fig:bitdestructionblocks}(b). In this section, a
$\widehat{G}_{\onebit}(t)$ will be obtained such that $D_{\onebit}$ scales
as $O(1/N)$. This is non-trivial to achieve because the \textit{non-linear}
quantization operation is coupled with the statistical estimation
procedure. The result will be established in two parts: (i) it will be
shown that suitable interpolation of one-bit samples converges to a
non-linear one-to-one function of $g(t)$ with an error term having a
pointwise variance of $O(1/N)$; and (ii) the obtained non-linear function
of $g(t)$ can be inverted in a stable manner  using recursive computation
based on contraction-mapping.  It will be assumed that $\var(W_d(t)) =
(\sigma^2 - \var(W(t)))_+$, where $\sigma$ is such that
(\ref{eq:variancecondition}) is satisfied.

The stability property of kernel $\phi(t)$ has been discussed
Sec.~\ref{sec:background}.  For this section, fix $\tau = 1/(N \lambda)$,
where $\lambda > 1$ is an arbitrary stability constant.  Analogous to
(\ref{eq:frameexpansion}), consider the random process obtained from the
single-bit samples $X(n\tau), n \in \zZ$,
\begin{align}
H_N(t) = \tau \sum_{n \in \zZ} (X(n \tau) - 1/2) \phi \left( t - n \tau
\right).\label{eq:intermediateestimate}
\end{align}
Then, the following proposition establishes the convergence of $H_N(t)$ to
a function of the signal of interest $g(t)$.
\begin{proposition}[Convergence of single-bit interpolation]
\label{prop:htconvergence}
Let $l(t) = (F(g(t)) - 1/2)$ and $H_N(t)$ be as defined in
(\ref{eq:intermediateestimate}). Then
\begin{align}
\sup_{t\in \rR} \eE(H_N(t) - l(t) \star \phi(t))^2 \leq \frac{C_2}{N} +
\frac{C_3}{N^2}, \label{eq:intermediateestimatelimit}
\end{align}
where $C_2>0$ and $C_3 > 0$ are constants independent of $N$.
\end{proposition}

\IEEEproof See Appendix~\ref{ap:onebitestimatoraccuracy}. \endproof

The factor $\tau = 1/(\lambda N)$ provides the normalization for averaging
in (\ref{eq:intermediateestimate}), while the terms $(X(n\tau) - 1/2)\phi(t
- n \tau)$ are weighted independent one-bit samples.  The average in
(\ref{eq:intermediateestimatelimit}) converges in mean-square to a
convolution.  The signal $l(t) \in \linf(\rR)$ and the limit $l(t) \star
\phi(t)$ is a lowpass version of $l(t)$.  The dependence of $l(t) \star
\phi(t)$ on $g(t)$ is non-linear due to quantization, which results in the
$F(g(t))$ term.  The original signal $g(t)$ is Zakai sense bandlimited and
it has \textit{one degree of freedom} per unit time.  The degree of freedom
per unit time of $l(t) \conv \phi(t)$ can be up to \textit{one} as well,
and $F(x)$ has `nice' properties as a function.  Thus, it is not
unreasonable to expect that there might be a class of $F(x)$ such that
$(F(g(t)) - 1/2) \conv \phi(t)$ can be inverted to find $g(t)$, even though
this equation is nonlinear.

Consider compandors defined by Landau and Miranker~\cite{landauMT1961}.
\begin{definition} \cite[pg~100]{landauMT1961}
A compandor is a monotonic function $Q(x)$ which has the property that
$Q(m(t)) \in \ltwo(\rR) $ if $m(t) \in \ltwo(\rR)$. 
\end{definition}
Landau and Miranker have shown that if $g(t) \in \ltwo(\rR)$ and
$\tilde{g}(\omega)$ is zero outside $[-\pi, \pi]$, and if $Q:[-1, 1]
\rightarrow \rR$ is a compandor with non-zero slope, then there is one to
one correspondence between $g(t)$ and $Q(g(t)) \star
\sinc(t)$~\cite{landauMT1961}.  Further, given any signal $m(t) \in
\ltwo(\rR)$ and $\tilde{m}(\omega)$ zero outside $[-\pi, \pi]$, there
exists a unique $g_m(t) \in \ltwo(\rR)$ with $\tilde{g_m}(\omega)$ zero
outside $[-\pi, \pi]$ and $Q(g_m(t)) \star \sinc(t) = m(t)$.

In our case, $g(t)$ need not be in $\ltwo(\rR)$, even though $l(t) =
F(g(t)) - 1/2$ is a compandor. Thus, the procedure of Landau and
Miranker does not extend directly to bandlimited signals in
$\linf(\rR)$, \textit{especially} in the presence of statistical
perturbations. Suitable modifications of their approach will be used
to obtain the results for our problem.

The dependence between $g(t)$ and $l(t) \star \phi(t)$ is quite non-linear.
There is no clear or obvious equation by which $g(t)$ can be obtained from
$l(t) \star \phi(t)$. Therefore, this inversion problem is casted into a
recursive setup, where Banach's fixed-point theorem can be leveraged along
with contraction mapping~\cite[Ch.~5]{kreyszigI1989}.  This approach is
inspired from the work of Landau and Miranker. Their recursive setup is
noted to be stable to perturbations of $g(t)$ in
$\ltwo(\rR)$~\cite{landauMT1961}.  This work will use a variant of their
procedure, since the perturbation due to statistical noise with finite
variance is not in $\ltwo(\rR)$.  Therefore, our recursive procedure to
obtain an estimate of $g(t)$ from $H_N(t)$ (see
(\ref{eq:intermediateestimatelimit})) and its analysis is non-trivial and
it will be presented in detail.

In summary, an estimate for $g(t)$ is required. Due to quantization and
noise, which is a non-linear operation, an approximation $H_N(t)$ of
$(F(g(t)) - 1/2) \star \phi(t)$ is available. The estimate $H_N(t)$, which
converges to $(F(g(t)) - 1/2) \star \phi(t)$ with sample density $N
\uparrow \infty$, will be inverted to obtain an estimate
$\widehat{G}_{\onebit}(t)$ for the signal $g(t)$.  To establish the
precision indifference principle, we wish to show that the mean-square
error $\sup_{t \in \rR} \eE|\widehat{G}_{\onebit}(t) - g(t)|^2$ decreases
as $O(1/N)$. The details are presented next.

A `clip to one' function $\clip[x]$ is defined first.
\begin{align}
\clip[x] & = x \quad \quad \quad \mbox{ if } |x| \leq 1 \nonumber \\
& = \mbox{sgn}(x) \quad \mbox{ otherwise}. \label{eq:clipto1}
\end{align}
Since $g(t)$ has a dynamic range bounded by one, by assumption, it will be
unaffected by clipping. Note that under the $\linf$ norm, this
transformation reduces the distance between any two scalars $x_1$ and
$x_2$, i.e., $|\clip[x_1] - \clip[x_2]| \leq |x_1 - x_2|$. This can be
verified on a case by case basis.  For example, if $x_1 > 1$ and $x_2 \in
[-1,1]$, then $|\clip[x_1] - \clip[x_2]| = |1 - x_2| \leq |x_1 - x_2|$.
Other cases can be similarly enumerated. This clipping procedure is
non-linear and complicates some of the presented analysis; however, we feel
that its presence is essential for analysis.

Let $\psi(t) = \phi(\lambda t)$. Then $\tilde{\psi}(\omega) =
\phi(\omega/\lambda)$. Thus, $\tilde{\psi}(\omega)$ is flat in
$[-\lambda \pi, \lambda \pi]$ and in $\pm[\lambda \pi, \lambda^2 \pi]$
decreases linearly to zero. Consider the set of bandlimited signals
defined by,
\begin{align}
\BLBounded = \{ m(t): |m(t)| \leq C_\phi \mbox{ and } m(t) \star
\psi(t) = m(t) \}. \label{eq:blbounded}
\end{align}
Then, $\BLBounded$ is a complete subset of the Banach space
$\linf(\rR)$.
\begin{lemma}[$\BLBounded$ is a complete metric space]
\label{lem:metric}
Let $\BLBounded$ be as defined in (\ref{eq:blbounded}). Then
$(\BLBounded, ||.||_\infty)$ is a complete subset of $(\linf(\rR),
||.||_\infty)$.
\end{lemma}

\IEEEproof Define the distance function $d: \BLBounded \times \BLBounded
\rightarrow \rR^+$ as $d(m_1, m_2) = ||m_1 - m_2||_\infty$ with
$m_1(t), m_2(t) \in \BLBounded$. It is easy to verify the axioms of
distance metric~\cite{kreyszigI1989}: (i) $d \geq 0$ and $d \leq
\infty$; (ii) $d(m_1, m_2) \equiv 0$ if and only if $m_1(t) = m_2(t)$;
(iii) $d(m_1, m_2) = d(m_2, m_1)$; and (iv) $d(m_1, m_2) \leq d(m_1,
m_3) + d(m_3, m_2)$ for any $m_1(t), m_2(t), m_3(t) \in \BLBounded$.

It is straightforward to see that $\BLBounded \subset \linf(\rR)$
since $||m||_\infty \leq C_\phi$ for  every $m(t) \in \BLBounded$.  To
show that the subset is complete, consider any Cauchy sequence $m_n(t)
\in \BLBounded$. Since $\linf(\rR)$ is complete, therefore $m_n(t)
\rightarrow s(t)$, where $s(t) \in \linf(\rR)$. It remains to show
that $s(t)$ belongs to $\BLBounded$. 

For any $\epsilon > 0$, there is an $n_0$ such that $||m_n -
s||_\infty < \epsilon$ for all $n > n_0$. Since $\int_{\rR} |\psi(t)|
\mbox{d}t = C_\phi/\lambda$, therefore, $||m_n \star \psi - s \star
\psi ||_\infty \leq ||m_n - s|| (C_\phi/\lambda) = C_\phi \epsilon
/\lambda$ for all $n > n_0$ (see Lemma~\ref{lemma:propagation}). Thus,
$m_n(t) \star \psi(t) \rightarrow s(t) \star \psi(t)$. However,
$m_n(t) \star \psi(t) \equiv m_n(t)$ since $m_n(t) \in \BLBounded$.
Therefore, it follows that $s(t) = s(t) \star \psi(t)$, or $s(t) \in
\BLBounded$. Thus, $\BLBounded$ is complete. \endproof

A map $T: \BLBounded \longrightarrow \BLBounded$ will be defined next.
This map will result in a recursive procedure to obtain $g(t)$ from
$h(t) := l(t) \conv \phi(t)$. Define
\begin{align}
T[m(t)] & = \clip \Big[ \mu h(t) + \left[ m(t) - \mu (F(m(t)) - 1/2) \right]
\star \phi(t) \Big]  \star \phi(t). \label{eq:keymap}
\end{align}
It will be shown that $T$ is a contraction on $(\BLBounded,
||.||_\infty)$.
\begin{lemma}[$T$ is a contraction]
\label{lem:tcontraction}
Let $(\BLBounded, ||.||_\infty)$ be the metric space as defined in
(\ref{eq:blbounded}). Let $T : \BLBounded \longrightarrow \BLBounded$
be a map as defined in (\ref{eq:keymap}). If the condition in
(\ref{eq:variancecondition}) is satisfied, then there is a choice of
$\mu$ such that $T$ is a contraction, i.e.,
\begin{align}
||T[m_1] - T[m_2] ||_\infty \leq \alpha ||m_1 - m_2||_\infty,
\end{align}
for some $0 < \alpha < 1$ and any $m_1(t), m_2(t) \in \BLBounded$. The
parameter $\alpha$ does not depend on the choice of $m_1$ and $m_2$.
\end{lemma}
\IEEEproof See Appendix~\ref{ap:lem_cont}. \endproof

Now the key recursive equation will be stated. Let $l(t) = (F(g(t)) -
1/2)$ and $h(t) = l(t) \star \phi(t)$ be available for obtaining
$g(t)$. Then,
\begin{align}
g_{k+1}(t) := T[g_{k}(t)]  = \clip \Big[ \mu h(t) + \left[ g_{k}(t) -
\mu (F(g_{k}(t)) - 1/2) \right] \star \phi(t) \Big] \star \phi(t),
\label{eq:recursion}
\end{align}
where $k \geq 0, k \in \zZ$ and $\mu > 0$ is a constant that will be
chosen according to Lemma~\ref{lem:tcontraction}. Set $g_0(t) \equiv
0$.  The original signal $g(t)$ is a fixed point of this equation and
it can be verified by substitution. The following proposition shows
that $g(t)$ is the \textit{only} fixed point of the equation in
(\ref{eq:recursion}). The proof hinges on Banach's fixed point theorem
or contraction theorem~\cite[Ch.~5]{kreyszigI1989}.
\begin{proposition}[Signal of interest is the fixed point of $T$]
\label{prop:fixedpt}
Let $g(t) \in \BLInterest \subset \BLBounded$ be a continuous bounded
bandlimited signal. Let $h(t) = l(t) \star \phi(t)$, where $l(t) = F(g(t))
- 1/2$. Consider the recursion $g_{k}(t) = T[g_{k-1}(t)]$, where $T$ is as
defined in (\ref{eq:keymap}). Set $g_0(t) \equiv 0$. If $\mu$ is selected
as in (\ref{eq:variancecondition}), then
\begin{align}
\lim_{k \rightarrow \infty } ||g_k - g||_\infty = 0. \label{eq:linfconv}
\end{align}
\end{proposition}
\IEEEproof  The proof is straightforward with Lemma~\ref{lem:metric} and
Lemma~\ref{lem:tcontraction} in place. Define $d(m_1, m_2) = ||m_1 -
m_2||_\infty$ for any $m_1(t), m_2(t) \in \BLBounded$. From
Lemma~\ref{lem:metric}, note that $(\BLBounded, d)$ is a complete metric
space.  The signal $g(t)$ is in $\BLBounded$ and it satisfies $g(t) =
T[g(t)]$, i.e., it is a fixed point for $T$ defined in (\ref{eq:keymap}).

Pick $\mu$ as in (\ref{eq:variancecondition}). Then $T$ is a contraction on
$(\BLBounded, d)$. Thus, by Banach's fixed point theorem (contraction
theorem)~\cite[Ch.~5]{kreyszigI1989}, there is \textit{exactly} one fixed
point in $\BLBounded$ for the equation $g(t) = T[g(t)]$. Since $g_k(t)$
converges to a fixed point, it must converge to $g(t)$ in the distance
metric $d$.  Thus the proof is complete.
\endproof

Propostion~\ref{prop:fixedpt} holds with perfect information about $l(t)
\star \phi(t)$. The estimation of signal from $H_N(t)$, the statistical
approximation of $l(t) \star \phi(t)$, will be discussed now.  Let
$G_{k}(t)$ be the sequence of random waveforms generated from $H_N(t)$ when
it is applied to the recursion in (\ref{eq:recursion}).  That is, fix
$G_0(t) \equiv 0$ and define 
\begin{align}
& G_{k+1}(t) := T[G_{k}(t)] =  \clip \Big[ \mu H_N(t) + \left[ G_{k}(t) -
\mu (F(G_{k}(t)) - 1/2) \right] \star \phi(t) \Big] \star \phi(t).
\label{eq:recursionG}
\end{align}
Let $\widehat{G}_{\onebit}(t) = \lim_{k \rightarrow \infty}
G_k(t)$.\footnote{This limit exists since it can be shown that $||G_k -
G_{k-1}||_\infty \leq \alpha ||G_{k-1} - G_{k-2}||_\infty$ for some $0  <
\alpha < 1$ by using an analogous procedure as in
Lemma~\ref{lem:tcontraction}.} For the same choice of $\mu$ which ensures
that $T$ is a contraction on $(\BLBounded, ||.||_\infty)$, the distortion
of $|\widehat{G}_{\onebit}(t) - g(t)|$ has to be established. To this end,
the following proposition is noted.
\begin{proposition}[1-bit estimation has distortion $O(1/N)$]
\label{prop:contraction_mse}
Let $H_N(t)$ be the estimate of $l(t)$ as described in
(\ref{eq:intermediateestimate}) and $\mu$ be selected as in
(\ref{eq:variancecondition}).  With $G_0(t) \equiv 0$, let $G_k(t)$ be the
sequence of random waveforms as  defined in (\ref{eq:recursionG}).  Define
$\lim_{k \rightarrow \infty} G_k(t) = \widehat{G}_{\onebit}(t)$. Then,
\begin{align}
D_{\onebit} := \sup_{t \in \rR} \eE( \widehat{G}_{\onebit}(t) - g(t))^2 =
O(1/N), \nonumber
\end{align}
i.e., the distortion $D_{\onebit}$ decreases as $O(1/N)$.
\end{proposition}

\IEEEproof See Appendix~\ref{ap:contraction_accuracy}. \endproof

The results of Proposition~\ref{prop:mse_unquantized} and
Proposition~\ref{prop:contraction_mse} can be summarized into the
following theorem.
\begin{theorem}[Precision indifference principle]
\label{thm:bitdestruction}
\textit{Let $g(t)$ be a bounded dynamic-range bandlimited-signal as
defined in (\ref{eq:blinterest}). Assume that $g(t) + W(t)$ is
available for sampling, where $W(t)$ an additive independent Gaussian
random process with finite variance. Fix an oversampling factor of
$N$, where $N$ is large for statistical averaging.  There exists an
estimate $\widehat{G}_{\onebit}(t)$ obtained from single-bit samples
of $g(t) + W(t)$ such that
\begin{align}
\sup_{t\in \rR} \eE| \widehat{G}_{\onebit}(t) - g(t) |^2 = O(1/N).
\nonumber
\end{align}
This distortion is proportional to the best possible distortion of
$O(1/N)$ that can be obtained with unquantized or perfect samples.}
\end{theorem}
A few remarks highlighting the importance of the results obtained will
conclude this section.

\subsection{Remarks on the results obtained}

\subsubsection{Comparison with the bit-conservation principle} The
bit-conservation principle~\cite{kumarIRH2011} is somewhat in contrast
to the precision indifference principle. Loosely speaking,
bit-conservation principle states that for sampling a bandlimited
signal in a noiseless setting, the oversampling density can be
traded-off against ADC precision while maintaining a fixed bit-rate
per Nyquist interval and an order-optimal pointwise distortion.  In
the presence of additive independent Gaussian noise, this tradeoff
between ADC precision and oversampling is absent while studying
pointwise mean-squared distortion. In the noisy setup, the distortion
is proportional to $1/N$, where $N$ is the oversampling density
irregardless of the ADC precision. The presence of noise shifts the
role of ADC precision towards only the proportionality constant in
distortion!

\subsubsection{Interpretation of precision-indifference principle}
First, it can be argued that the precision indifference principle
holds while estimating a constant signal (one degree of freedom) in
additive independent Gaussian noise.  Assume that a constant $c \in
[-1,1]$ has to be estimated based on $N$ noisy readings $Y_i = c+W_i,
1 \leq i \leq N$, where $\{W_i, 1 \leq i \leq N\}$ are i.i.d.~${\cal
N}(0, \sigma^2)$.  In the absence of quantization, $\widehat{C}_N =
(\sum_{i = 1}^N Y_i)/N$  converges to $c$ in the mean-square sense,
and $\eE(\widehat{C}_N - c)^2 = \sigma^2/N$. This is the optimal
distortion if perfect (unquantized) samples are available. Now
consider the case where single-bit readings $B_i = \indicator(c + W_i
\geq 0), 1 \leq i \leq N$ are available.  The random variables $\{B_i,
1 \leq i \leq N\}$ are i.i.d.~$\mbox{Ber}(q)$ where $q = \pP(W \geq -
c) = \pP(W \leq c) = F(c)$. Assume $\widehat{B}_N = (\sum_{i = 1}^N
B_i)/N$. It can be shown that $\eE(\widehat{B}_N - F(c))^2 \leq
1/(4N)$ since each $\mbox{var}(B_i) \leq F(c)(1-F(c)) \leq 1/4$.
Define $\widehat{C}_{\onebit} = F^{-1}(\widehat{B}_N)$ if
$\widehat{B}_N \in [F(-1), F(1)]$ and $\widehat{C}_{\onebit} = \pm 1$
otherwise.  Since $F(x)$ is invertible and
$\mbox{d}F^{-1}(x)/\mbox{d}x$ is bounded for $x \in [F(-1),F(1)]$,
therefore, using the delta method, $\widehat{C}_{\onebit}$ obtained
from $\widehat{B}_N$ has a mean-squared error which decreases as
$(1/N)$~\cite{bickelDM2001}.  Next, it should be noted that
bandlimited signals have one degree of freedom in every Nyquist
interval. An oversampling factor of $N$ means that there are $N$
samples to observe each degree of freedom  on an average. Finally,
observing the Nyquist samples of a bandlimited signals with a
distortion of $O(1/N)$ results, by stable interpolation with kernel
$\phi(t)$, in a pointwise distortion of $O(1/N)$ for the signal
estimate at any point.

\subsubsection{Precision-indifference for a larger class of noise}
Consider the model where each sample $Y(n\tau) = g(n\tau) + V(n\tau)$ is
affected by some non-Gaussian noise.  Focus on the case where $V(n\tau)$
can be written as $V(n\tau) = W(n\tau) + U(n\tau)$, where $W(n\tau)$ and
$U(n\tau)$ are i.i.d.~for all $n \in \zZ, \tau \in \rR$. If $W(n\tau)$ is
Gaussian, $\var(V(n\tau)) = \sigma^2 < \infty$, and $F_V(x)$ satisfies
(\ref{eq:variancecondition}),  then the precision indifference principle
will hold.  The extension of existing proofs is simple and only its key
steps will be mentioned here. In the perfect sample case (see
Fig.~\ref{fig:bitdestructionblocks}(a)), the Gaussian part of $V(t)$ will
limit the best possible (optimal) distortion to $O(1/N)$; this is because
even if the values of $U(n\tau)$ are (magically) known the residual
$W(n\tau)$ will limit the distortion.  With single-bit quantization, note
that all the proofs in Sec.~\ref{sec:onebit} only depend upon the existence
of a $\delta$ and $\Delta$ such that (\ref{eq:variancecondition}) is
satisfied, monotonicity of $F_V(x)$ such that its derivative is bounded
away from zero, and $F_V(0) = 1/2$.  The recursive procedure in
(\ref{eq:recursion}), however, requires the knowledge of $F_V(x)$.

\section{Conclusions and future work}
\label{sec:conclusions}

The sampling, quantization, and estimation of a bounded dynamic-range
bandlimited signal affected by additive independent Gaussian noise was
studied.  Such setup naturally arises in distributed sampling or where the
sampling device itself is noisy. For bandlimited signals, the distortion
due to additive independent Gaussian noise can be reduced by oversampling
(statistical diversity).  The maximum pointwise expected mean-squared error
(statistical $\ltwo$ error) was used as a distortion metric.  Using two
extreme scenarios of quantizer precision, namely infinite precision and
single-bit precision, a quantizer precision indifference principle was
illustrated. It was shown that the optimal law for distortion is $O(1/N)$,
where $N$ is the oversampling ratio with respect to the Nyquist rate. This
scaling of distortion is unaffected by the \textit{quantizer precision},
which is the key message of the precision indifference principle.  In other
words, the reconstruction distortion law, up to a proportionality constant,
is unaffected by quantizer precision.

Extensions of the precision indifference principle to other classes of
parametric or non-parametric signals is of immediate interest. Further,
this work assumed sufficient dithering by noise because the estimators were
linear.  It is of interest to look towards estimation techniques which do
not require extra dithering.

\section*{Acknowledgment}

The problem of sampling a smooth signal in the presence of noise in a
distributed setup was suggested by Prof. Kannan Ramchandran, EECS,
University of California, Berkeley, CA. Discussions on this problem
with Prof. Kannan Ramchandran and Prof.~Martin Wainwright, EECS,
University of California, Berkeley, CA, Prof. H.~Narayanan, EE, IIT
Bombay, and Prof. Prakash Ishwar, ECE, Boston University, Cambridge,
MA were insightful.

\bibliographystyle{IEEEtran}

\begin{thebibliography}{10}
\providecommand{\url}[1]{#1}
\csname url@samestyle\endcsname
\providecommand{\newblock}{\relax}
\providecommand{\bibinfo}[2]{#2}
\providecommand{\BIBentrySTDinterwordspacing}{\spaceskip=0pt\relax}
\providecommand{\BIBentryALTinterwordstretchfactor}{4}
\providecommand{\BIBentryALTinterwordspacing}{\spaceskip=\fontdimen2\font plus
\BIBentryALTinterwordstretchfactor\fontdimen3\font minus
  \fontdimen4\font\relax}
\providecommand{\BIBforeignlanguage}[2]{{%
\expandafter\ifx\csname l@#1\endcsname\relax
\typeout{** WARNING: IEEEtran.bst: No hyphenation pattern has been}%
\typeout{** loaded for the language `#1'. Using the pattern for}%
\typeout{** the default language instead.}%
\else
\language=\csname l@#1\endcsname
\fi
#2}}
\providecommand{\BIBdecl}{\relax}
\BIBdecl

\bibitem{mallatSA2009}
S.~Mallat, \emph{A Wavelet Tour of Signal Processing: The Sparse Way}.\hskip
  1em plus 0.5em minus 0.4em\relax Burlington, MA, USA: Academic Press, 2009.

\bibitem{pinskerO1980}
{M.~S.~Pinsker}, ``Optimal filtration of square-integrable signals in
  {Gaussian} noise,'' \emph{Problemy Peredachi Informatsil}, vol.~16, no.~2,
  pp. 52--68, Apr. 1980.

\bibitem{grayo1987}
R.~M. Gray, ``{Oversampled Sigma-Delta Modulation},'' \emph{IEEE Trans.
  Commun.}, vol.~35, no.~5, pp. 481--489, May 1987.

\bibitem{zoranDLS2007}
Z.~Cvetkovi\'{c}, I.~Daubechies, and B.~Logan, ``{Single-Bit Oversampled A/D
  Conversion With Exponential Accuracy in the Bit Rate},'' \emph{IEEE
  Transactions on Information Theory}, vol.~53, no.~11, pp. 3979--3989, Nov.
  2007.

\bibitem{kumarIRH2011}
A.~Kumar, P.~Ishwar, and K.~Ramchandran, ``High-resolution distributed sampling
  of bandlimited fields with low-precision sensors,'' \emph{IEEE Transactions
  on Information Theory}, vol.~57, no.~1, pp. 476--492, Jan. 2011.

\bibitem{bickelDM2001}
P.~J. Bickel and K.~A. Doksum, \emph{Mathematical Statistics Vol I}.\hskip 1em
  plus 0.5em minus 0.4em\relax Upper Saddle River, NJ, USA: Prentice Hall,
  2001.

\bibitem{GrayNIT98}
{R.~M.~Gray and D.~L.~Neuhoff}, ``Quantization,'' \emph{IEEE Transactions on
  Information Theory}, vol. {IT-44}, pp. 2325--2383, {Oct.} 1998.

\bibitem{masryT1981}
E.~Masry, ``The reconstruction of analog signals from the sign of their noisy
  samples,'' \emph{IEEE Transactions on Information Theory}, vol.~27, no.~6,
  pp. 735--745, Nov. 1981.

\bibitem{wangID2009}
Y.~Wang and P.~Ishwar, ``Distributed field estimation with randomly deployed,
  noisy, binary sensors,'' \emph{{IEEE T}ransactions on Signal Processing},
  vol.~57, no.~3, pp. 1177--1189, Mar. 2009.

\bibitem{masryIF2009}
E.~Masry and P.~Ishwar, ``Field estimation from randomly located binary noisy
  sensors,'' \emph{{IEEE T}ransactions on Information Theory}, vol.~55, no.~11,
  pp. 5197--5210, Nov. 2009.

\bibitem{zakaiB1965}
M.~Zakai, ``Band-limited functions and the sampling theorem,''
  \emph{Information and Control}, vol.~8, pp. 143--158, 1965.

\bibitem{cambanisMZ1976}
S.~Cambanis and E.~Masry, ``Zakai's class of bandlimited functions and
  processes: Its characterization and properties,'' \emph{{SIAM} Journal on
  Applied Mathematics}, vol.~30, no.~1, pp. 10--21, Jan. 1976.

\bibitem{gershogray}
A.~Gersho and R.~M. Gray, \emph{Vector Quantization and Signal
  Compression}.\hskip 1em plus 0.5em minus 0.4em\relax Boston: Kluwer Academic,
  1992.

\bibitem{slepianO1976}
D.~Slepian, ``On bandwidth,'' \emph{Proceedings of the IEEE}, vol.~64, no.~3,
  pp. 292--300, Mar. 1976.

\bibitem{landauMT1961}
H.~J. Landau and W.~L. Miranker, ``The recovery of distorted band-limitd
  signals,'' \emph{Journal of Mathematical Analysis and Applications}, vol.~2,
  no.~1, pp. 97--104, Feb. 1961.

\bibitem{kreyszigI1989}
E.~Kreyszig, \emph{Introductory Functional Analysis with Applications},
  1st~ed.\hskip 1em plus 0.5em minus 0.4em\relax Wiley, 1989.

\bibitem{pelgromA2010}
M.~J.~M. Pelgrom, \emph{Analog-to-Digital Conversion}.\hskip 1em plus 0.5em
  minus 0.4em\relax Springer, NY, 2010.

\bibitem{rudinp1976}
W.~Rudin, \emph{{Principles of Mathematical Analysis}}.\hskip 1em plus 0.5em
  minus 0.4em\relax USA: McGraw-Hill Companies, 1976.

\end{thebibliography}


\section{Appendix}

\subsection{Unquantized samples result in a distortion of $O(1/N)$}
\label{ap:unquantizedMSE}

By using (\ref{eq:frameexpansion}) and the definition of
$\widehat{G}_{\frame}(t)$, first note that,
\begin{eqnarray}
\widehat{G}_{\frame}(t) - g(t) = \frac{1}{N} \sum_{i = 0}^{N-1} \lambda \sum_{k \in
\zZ} W\left( \frac{k}{\lambda} + i \tau \right) \phi\left(t -
\frac{k}{\lambda} - \frac{i}{N \lambda}\right). \nonumber
\end{eqnarray}
Since $W(k + i \tau)$ are i.i.d.~with variance $\sigma^2$, therefore,
\begin{eqnarray}
\eE|\widehat{G}_{\frame}(t) - g(t)|^2 = \frac{1}{N^2} \sum_{i = 0}^{N-1}
\sum_{k \in \zZ} \lambda^2 \sigma^2   \left| \phi\left(t - \frac{k}{\lambda} -
\frac{i}{N \lambda}\right)\right|^2. \nonumber % \label{eq:temp1}
\end{eqnarray}
From (\ref{eq:squaresummablephi}), $\sum_{k \in \zZ} |\phi(t -
\frac{k}{\lambda})|^2 \leq C_{\phi}''$. Thus,
\begin{align}
\eE|\widehat{G}_{\frame}(t) - g(t)|^2 \leq \frac{C_{\phi}'' \lambda^2 \sigma^2}{N}.
\label{eq:gframebound}
\end{align}
The proof is now complete.

\subsection{Estimation of non-linear function of the signal $g(t)$}
\label{ap:onebitestimatoraccuracy}

By the linearity of expectation, it is easy to see that $\eE(H_N(t)) = \tau
\sum_{n \in \zZ} l(n \tau) \phi \left( t - n \tau \right)$.\footnote{This
summation is well defined since $l(t) = (F(g(t)) - 1/2)$ is bounded and
$\phi(t)$ is integrable.} For variance calculations, note that each $X(n
\tau)$ is an indicator random variable. Thus, $\var(X(n\tau)) \leq (1/4)$.
Using the independence of $\{X(n\tau), n \in \zZ\}$, we get
\begin{eqnarray}
\var(H_N(t)) & = & \tau^2 \sum_{n \in \zZ} \var(X(n\tau)) \left| \phi
\left( t - n \tau \right)\right|^2, \nonumber \\
& \leq & \frac{\tau^2}{4} \sum_{l \in \zZ} \sum_{i = 0}^{N-1} \left| \phi
\left( t - \frac{l}{\lambda} - i \tau \right)\right|^2, \nonumber  \\
& \leq & \frac{\tau^2}{4}  N C_{\phi}'' = \frac{C_{\phi}''}{4\lambda^2 N} =
\frac{C_2}{N}, \nonumber
\end{eqnarray}
where $C_{\phi}''$ is given by (\ref{eq:squaresummablephi}), and is finite
since $\phi (t)$ decays rapidly and is in $\ltwo(\rR)$. The constant $C_2
:= C_{\phi}''/(4 \lambda^2)$.

Next it will be shown that as $N \rightarrow \infty$, the expression $\tau
\sum_{n \in \zZ} F(g(n\tau)) \phi  \left( t - n \tau \right)$ converges to
the right hand side of (\ref{eq:intermediateestimatelimit}). The proof uses
the fast decay of $\phi(t)$, which ensures absolute integrability of $\phi
(t)$.  Consider the expression
\begin{align}
e_0 = \left[\int_{0}^{\tau} l(u) \phi(t - u) \mbox{d} u \right] - \tau
l(\tau) \phi(t - \tau). \nonumber 
\end{align}
This expression represents the error in approximating convolution integral
in a $\tau$ interval by the corresponding term in a Riemann
sum~\cite{rudinp1976}.  This can be bounded as explained next,
\begin{align}
|e_0| & = \left| \int_{0}^{\tau} [l(u) - l(\tau)] \phi(t - u) \mbox{d}u -
l(\tau) \left[\tau \phi(t - \tau) - \int_{0}^{\tau} \phi(t - u)
\mbox{d}u\right] \right|,  \nonumber\\
& \leq ||l'||_\infty \tau \int_{0}^{\tau} |\phi(t-u)| \mbox{d} u + \left|
l(\tau) \left[\tau \phi(t - \tau) - \int_{0}^{\tau} \phi(t - u)
\mbox{d}u\right]\right|. \label{eq:Htconv1}
\end{align}
where we use the triangle inequality, $|l(u) - l(\tau)| \leq
||l'||_\infty(u - \tau)$, and finally $|u - \tau| \leq \tau$. Next, by
Lagrange mean-value theorem, we note that $\int_{0}^{\tau} \phi(t - u)
\mbox{d} u = \tau \phi(t - t_0)$ for some $t_0 \in (0, \tau)$. Thus, from
(\ref{eq:Htconv1}),
\begin{align}
|e_0| & \leq ||l'||_\infty \tau \int_{0}^{\tau} |\phi(t-u)| \mbox{d} u +
||l||_\infty \tau |\phi(t - \tau) - \phi(t - t_0)|, \nonumber \\
& \leq ||l'||_\infty \tau \int_{0}^{\tau} |\phi(t - u)|\mbox{d}u +
||l||_\infty \tau^2 |\phi'(t - u_0)|, \nonumber
\end{align}
for some $u_0 \in (0, \tau)$. We note that $|\phi(t - u) - \phi(t - t_0)| =
|\phi'(t - u_0)| |u - t_0| \leq $ for some $u_0 \in (t_0, \tau) \subseteq
(0, \tau)$ (this is by Lagrange's mean-value theorem).  In the same
fashion, for any $[n \tau, (n+1)\tau]$ interval, the following bound can be
established:
\begin{align}
|e_n| & \leq ||l'||_\infty \tau \int_{n\tau}^{(n+1)\tau} |\phi(t -
u)|\mbox{d}u + ||l||_\infty \tau^2 |\phi'(t - u_n)|, \label{eq:Htconv2}
\end{align}
where $u_n \in (n\tau, (n+1)\tau)$. Finally,
\begin{align}
|\eE(H_N(t)) - l(t) \star \phi(t)| & = \left| \tau \sum_{n\in \zZ} l(n\tau)
\phi(t - n \tau) - \int_{u \in \rR} l(u) \phi(t - u) \mbox{d} u\right|,
\nonumber \\
& \leq \left| \sum_{n \in \zZ} e_n \right|, \nonumber \\
& \leq \sum_{n \in \zZ} |e_n|, \nonumber  \\
& \leq ||l'||_\infty \tau \int_{u \in \rR} |\phi (t - u)|\mbox{d} u +
||l||_\infty \tau^2 \left[ \sum_{n\in \zZ} |\phi (t - u_n)|\right],
\nonumber 
\end{align}
where the last step follows using (\ref{eq:Htconv2}). Using the boundedness
of $C_{\phi}$ and $C_{\phi}'$ (see (\ref{eq:integrablephi}) and
(\ref{eq:summablephiprime})), and using bounds for $l(t)$ and $l'(t)$, we
get,
\begin{align}
|\eE(H_N(t)) - l(t) \star \phi(t))| \leq \left( \frac{ F'(0) (2\pi^2)
C_{\phi}}{\lambda} + \frac{|F(1) - 1/2| C_{\phi}'}{\lambda^2} \right)
\frac{1}{N} = \frac{C_3}{N}.  \label{eq:Htaccuracy}
\end{align}
Finally,
\begin{align}
\eE(H_N(t) - l(t) \star \phi(t))^2  & = \var(H_N(t)) + [\eE(H_N(t)) - l(t)
\star \phi(t)]^2 \nonumber \\
& \leq \frac{C_2}{N} + \frac{C_3^2}{N^2} \rightarrow 0 \mbox{ as } N
\rightarrow \infty. \nonumber
\end{align}
The upper bound in the above inequality is independent of $t$; therefore,
the desired result follows by maximizing the left hand side as a function
of $t$.  This completes the proof. It should be noted that the mean-squared
error between $H_N(t)$ and $l(t) \star \phi(t)$ is of the order of $(1/N)$.
This result will be used to find the mean-squared error of
$\widehat{G}_{\onebit}(t) - g(t)$.

\subsection{The map $T$ is a contraction}
\label{ap:lem_cont}

From (\ref{eq:variancecondition}), first note
\begin{align}
\left( 1 - \frac{1}{C^2_\phi}\right) \frac{1}{\delta} < \left( 1 -
\frac{1}{\sqrt{2} C^2_\phi} \right) \frac{1}{\delta} < \mu <
\frac{1}{\Delta}.  \nonumber
\end{align}
Since $F'(x)$, the pdf of the noise random variable, is positive and $F'(x)
\in [\delta, \Delta]$ for $x \in [-C_\phi,C_\phi]$, therefore,
\begin{align}
\delta \leq \frac{F(x) - F(y)}{x - y} \leq \Delta \mbox{ for } x \neq y
\mbox{ and } x, y \in [-C_\phi,C_\phi]. \label{eq:cdfsmoothness}
\end{align}
Now consider the map in (\ref{eq:keymap}). First, it will be shown that if
$m(t) \in \BLBounded$, then $T[m(t)] \in \BLBounded$. Define,
\begin{align}
r(t) := \mu h(t) + [m(t) - \mu (F(m(t)) - 1/2)] \star \phi(t).
\label{eq:rt}
\end{align}
Then $T[m(t)] = \clip[r(t)] \star \phi(t)$.  Since $|\clip[r(t)]| \leq 1$,
Lemma~\ref{lemma:propagation} applied with $p(t) = \clip[r(t)]$ results in
$|\clip[r(t)] \star \phi(t)| \leq C_\phi$ for all $t\in \rR$. Next,
$T[m(t)] \star \psi(t) = \left(\clip[r(t)] \star \phi(t)\right) \star
\psi(t) = \clip[r(t)] \star \left( \phi(t) \star \psi(t) \right) =
\clip[r(t)] \star \phi(t) = T[m(t)]$. Note that $\tilde{\phi}(\omega)
\tilde{\psi}(\omega) = \tilde{\phi}(\omega)$ which results in $\phi(t)
\star \psi(t) = \phi(t)$. That is, $T[m(t)]$ satisfies the convolution
property needed to be present in the set $\BLBounded$.

The contraction property will now be established.  Let $m_1(t)$ and
$m_2(t)$ be any two signals in $\BLBounded$ with corresponding
$r_1(t)$ and $r_2(t)$ as defined in (\ref{eq:rt}). The transformed
signals can be written in terms of $r_i(t)$ as $T[m_i(t)] =
\clip[r_i(t)] \star \phi(t)$ for $i  = 1,2$.  The signals $m_1(t)$ and
$m_2(t)$  are bounded in $[-C_\phi,C_\phi]$. The contraction property
is established by the following steps:
\begin{align}
|r_1(t) - r_2(t)| & = \left| \left[ m_1(t) - m_2(t) - \mu \left(
F(m_1(t)) - F(m_2(t)) \right) \right] \star \phi(t) \right| \nonumber
\\
& = \left| \int_{u \in \rR} \left( m_1(u) - m_2(u) \right) \left(1 -
\mu \frac{F(m_1(u)) - F(m_2(u))}{m_1(u) - m_2(u)} \right) \phi(t - u)
\mbox{d}u\right| \nonumber \\
& \stackrel{(a)}{\leq}  \int_{u \in \rR} \left| m_1(u) - m_2(u)
\right| \left|1 -  \mu \frac{F(m_1(u)) - F(m_2(u))}{m_1(u) - m_2(u)}
\right| |\phi(t - u)| \mbox{d}u \nonumber  \\
& \stackrel{(b)}{\leq} \int_{u \in \rR} \left| m_1(u) - m_2(u) \right|
|1 - \mu \delta| |\phi(t-u)| \mbox{d} u  \label{eq:deltacondition1} \\ 
& \stackrel{(c)}{\leq} |1 - \mu \delta| ||m_1 - m_2||_\infty \int_{u
\in \rR} |\phi(t-u)| \mbox{d} u  \nonumber  \\
& = C_{\phi} |1 - \mu \delta | \cdot ||m_1 - m_2||_\infty, \nonumber 
\end{align}
where $(a)$ follows by the triangle inequality, $(b)$ follows from
(\ref{eq:cdfsmoothness}) and $\mu \Delta < 1$, and $(c)$ follows from
the definition of the $\linf$-norm. Next, by using the
distance-reduction property of the clip-to-one function, we get
$|\clip[r_1(t)] - \clip[r_2(t)]| \leq |r_1(t) - r_2(t)| \leq C_{\phi}
|1 - \mu \delta | \cdot ||m_1 - m_2||_\infty$.  Applying
Lemma~\ref{lemma:propagation} with $p(t) = r_1(t) - r_2(t)$, we get,
\begin{align}
|T[m_1(t)] - T[m_2(t)]| & \leq C^2_{\phi} |1 - \mu \delta | \cdot
||m_1 - m_2||_\infty. \nonumber 
\end{align}
By taking supremum on $t$ in the left hand side of the above equation,
the desired contraction can be obtained:
\begin{align}
||T[m_1] - T[m_2]||_\infty \leq C^2_{\phi} |1 - \mu \delta| \cdot
||m_1 - m_2||_\infty, \label{eq:tcontraction}
\end{align}
independent of the choice of $m_1(t), m_2(t) \in \BLBounded$. The
conditions in (\ref{eq:variancecondition}) ensure that the parameter
$C^2_{\phi} | 1 - \mu \delta| < 1$. Set $\alpha := C^2_{\phi} | 1 -
\mu \delta|$, where $\alpha < 1$.  Thus,
\begin{align}
||T[m_1] - T[m_2]||_\infty \leq \alpha \cdot ||m_1 - m_2||_\infty,
\nonumber 
\end{align}
for some $0 < \alpha < 1$. Since $C_\phi$, $\delta$, and $\Delta$ do
not depend on $m_1$ and $m_2$, therefore $\alpha$ is independent of
the choice of $m_1$ and $m_2$. Thus the proof is complete.

\subsection{Mean-squared error analysis of $|\widehat{G}_{\onebit}(t) -
g(t)|$ using contraction}
\label{ap:contraction_accuracy}

To analyze the mean-squared error, two sets of recursion will be
considered.  One will involve $H_N(t)$, the statistical estimate of
$l(t) \star \phi(t)$, and corresponding $\widehat{G}_k(t)$. Then
$\widehat{G}_{\onebit}(t)$ is the limit of $\widehat{G}_k(t)$ as $k
\rightarrow \infty$.  The second recursion will involve $h(t) =
l(t)\star\phi(t)$ and the corresponding estimate $g_k(t)$ for the
bandlimited signal $g(t)$. In the second recursion, $g(t)$ is the
limit of $g_k(t)$. Let $r_k(t)$ and $R_k(t)$ be defined as follows:
\begin{align}
r_k(t) & = \mu h(t) - \left[ g_{k-1}(t) - \mu (F(g_{k-1}(t)) - 1/2)
\right] \star \phi(t), \label{eq:rk} \\
R_k(t) & = \mu H_N(t) - \left[ G_{k-1}(t) - \mu (F(G_{k-1}(t)) - 1/2)
\right] \star \phi(t). \label{eq:Rk}
\end{align}
Note that $g_k(t) = \clip[r_k(t)] \star \phi(t)$ and $G_k(t) =
\clip[R_k(t)] \star \phi(t)$.  By subtracting (\ref{eq:rk}) from
(\ref{eq:Rk}), the following equations are obtained:
\begin{align}
R_k(t) - r_k(t) & = \mu(H_N(t) - h(t)) - \Big[G_{k-1}(t) - g_{k-1}(t)
-  \mu (F(G_{k-1}(t)) - F(g_{k-1}(t)) \Big] \star \phi(t), \nonumber
\\
& = \mu(H_N(t) - h(t)) - \int_{u \in \rR} \phi(t - u) \Big[G_{k-1}(u)
- g_{k-1}(u) -  \mu(F(G_{k-1}(u)) - F(g_{k-1}(u)) ) \Big] \mbox{d}u.
\nonumber
\end{align}
By applying the triangle inequality twice on the above equation, the
following inequalities are obtained: 
\begin{align}
|R_k(t) - r_k(t)| & \leq \mu |H_N(t) - h(t)| + \Bigg| \int_{u \in \rR}
\phi(t - u) \Big[ G_{k-1}(u)  - g_{k-1}(u) -  \mu(F(G_{k-1}(u)) -
F(g_{k-1}(u)) ) \Big] \mbox{d}u \Bigg| \nonumber \\
& \leq \mu |H_N(t) - h(t)| + \int_{u \in \rR} |\phi(t - u)|
\left|G_{k-1}(u)  - g_{k-1}(u)\right|  \left| 1  - \mu
\frac{F(G_{k-1}(u)) - F(g_{k-1}(u))}{G_{k-1}(u) - g_{k-1}(u)}\right|
\mbox{d}u \label{eq:g1kerror}
\end{align}
The use of (\ref{eq:variancecondition}) and (\ref{eq:cdfsmoothness})
in (\ref{eq:g1kerror}) results in
\begin{align}
|R_k(t) - r_k(t)| & \leq \mu |H_N(t) - h(t)| + |1 - \mu \delta|
\int_{u \in \rR} |\phi(t-u)| |G_{k-1}(u) - g_{k-1}(u)| \mbox{d}u.
\nonumber
\end{align}
Since the clipping operation reduces distance, therefore
\begin{align}
|\clip[R_k(t)] - \clip[r_k(t)] | & \leq \mu |H_N(t) - h(t)| + |1 - \mu
\delta| \int_{u \in \rR} |\phi(t-u)| |G_{k-1}(u) - g_{k-1}(u)|
\mbox{d}u. \nonumber \\
& = \mu |H_N(t) - h(t)| + |1 - \mu \delta| \left( |\phi(t)| \star
|G_{k-1}(t) - g_{k-1}(t)| \right). \label{eq:msecoreeqn}
\end{align}
Now the mean-squared error of $\clip[R_k(t)] - \clip[r_k(t)]$ will be
bounded using (\ref{eq:msecoreeqn}). First note that for any two
random variables $X$ and $Y$, $\eE((X+Y)^2) \leq 2 \eE(X^2 + Y^2)$.
Thus, taking second moments on both sides of (\ref{eq:msecoreeqn})
results in
\begin{align}
\eE(|\clip[R_k(t)] - \clip[r_k(t)]|^2) & \leq  2 \mu^2 \eE(|H_N(t) -
h(t)|)^2 + 2 |1 - \mu \delta|^2 \eE\left( (|\phi(t)| \star |G_{k-1}(t)
- g_{k-1}(t)|)^2 \right), \nonumber \\
\leq & 2 \mu^2 \eE|H_N(t) - h(t)|^2 +  \quad 2 |1 - \mu \delta|^2
\left[ \sup_{t} \eE|G_{k-1}(t) - g_{k-1}(t)|^2 \right] C_{\phi}^2,
\end{align}
where the last inequality follows from Lemma~\ref{lemma:propagation}
with $P(t) = G_{k-1}(t) - g_{k-1}(t)$. Taking supremum over $t$ on the
left side, the following recursive relationship is obtained:
\begin{align}
& \sup_{t} \eE(|\clip[R_k(t)] - \clip[r_k(t)]|)^2 \leq 2 \mu^2 \left(
\frac{C_2}{4 \lambda^2 N} + \frac{C_3}{N^2} \right) +  2 C_{\phi}^2 |1
- \mu \delta|^2 \sup_{t} \eE(|G_{k-1}(t) - g_{k-1}(t)|)^2,
\label{eq:clipMSE}
\end{align}
where the uniform upper bound on $\eE|H_N(t) - h(t)|^2$ from
(\ref{eq:intermediateestimatelimit}) has been used. Since $G_k(t) -
g_k(t) =  (\clip[R_k(t)] - \clip[r_k(t)] ) \star \phi(t)$, by applying
Lemma~\ref{lemma:propagation} with $P(t) = \clip[R_k(t)] -
\clip[r_k(t)]$ we get,
\begin{align}
\sup_{t} \eE(|G_k(t) - g_k(t)|)^2 \leq 2 C^2_\phi \mu^2 \left(
\frac{C_2}{4 \lambda^2 N} + \frac{C_3}{N^2} \right) + 2 C_{\phi}^4 |1
- \mu \delta|^2 \sup_{t} \eE(|G_{k-1}(t) - g_{k-1}(t)|)^2.
\label{eq:supMSE}
\end{align}
From (\ref{eq:variancecondition}), it is noted that $\sqrt{2}
C^2_\phi |1 - \mu \delta| < 1$. Define $0 < \beta := 2 C_{\phi}^4 |1 -
\mu \delta|^2 < 1$. Using the recursion in (\ref{eq:supMSE}), it
follows that
\begin{align}
\lim_{k \rightarrow \infty} \sup_{t} \eE|G_k(t) - g_k(t)|^2 \leq
\frac{1}{1-\beta} 2 C^2_\phi \mu^2 \left( \frac{C_2}{4 \lambda^2 N} +
\frac{C_3}{N^2} \right)
\end{align}
Since we know that $G_k(t)$ and $g_k(t)$ converge in $\linf(\rR)$ to
$\widehat{G}_{\onebit}(t)$ and $g(t)$, respectively, therefore,
\begin{align}
\sup_{t} \eE|\widehat{G}_{\onebit}(t) - g(t)|^2 \leq \frac{1}{1-\beta}
2 \mu^2 \left( \frac{C_2}{4 \lambda^2 N} + \frac{C_3}{N^2} \right)
\end{align}
Lastly, $\mu, \beta, C_2, \lambda$, and $C_3$ are constants that do not
depend on $N$, therefore,
\begin{align}
\sup_{t} \eE|\widehat{G}_{\onebit}(t) - g(t)|^2 = O(1/N).\nonumber
\end{align}
The proportionality constant depends upon the class of signal
$\BLBounded$, the noise variance $\sigma^2$, the stability properties
of $\phi(t)$, and the chosen constant $\mu$. It does not depend on the
individual signal $g(t)$. This completes the proof of the accuracy
indifference principle.

\end{document}